\documentclass[prb,superscriptaddress,twocolumn,reprint]{revtex4-1}

\usepackage{amssymb}
\usepackage{amsfonts}
\usepackage{amsmath}
\usepackage{graphicx}
\usepackage{xcolor}
\usepackage{dcolumn}
\usepackage{bm}
\usepackage{booktabs}   
\usepackage{multirow}   
\usepackage{xfrac}      
\usepackage{braket}     
\usepackage{textcomp}
\usepackage{epstopdf}
\usepackage{wasysym}
\usepackage[colorlinks]{hyperref}
\hypersetup{
                pdfstartview={FitH},
                linkcolor=blue,
                citecolor=blue,
                filecolor=blue,
                urlcolor=blue
}

\newcommand\varpm{\mathbin{\vcenter{\hbox{
  \oalign{\hfil$\scriptstyle+$\hfil\cr
          \noalign{\kern-.3ex}
          $\scriptscriptstyle({-})$\cr}
}}}}
\newcommand\varmp{\mathbin{\vcenter{\hbox{
  \oalign{\hfil$\scriptscriptstyle-$\hfil\cr
          \noalign{\kern-.3ex}
          $\scriptstyle({+})$\cr}
}}}}

\begin{document}

\title{Optically induced orbital polarization in bulk germanium}


\author{F. Scali} 
\affiliation{Dipartimento di Fisica, Politecnico di Milano, Piazza Leonardo da Vinci 32, Milan 20133, Italy}

\author{M. Finazzi}
\affiliation{Dipartimento di Fisica, Politecnico di Milano, Piazza Leonardo da Vinci 32, Milan 20133, Italy}

\author{F. Bottegoni}
\affiliation{Dipartimento di Fisica, Politecnico di Milano, Piazza Leonardo da Vinci 32, Milan 20133, Italy}

\author{C. Zucchetti}
\email{carlo.zucchetti@polimi.it}
\affiliation{Dipartimento di Fisica, Politecnico di Milano, Piazza Leonardo da Vinci 32, Milan 20133, Italy} 

\date{\today}


\begin{abstract}

Optical orientation has been proven as a powerful tool to inject spin-polarized electron and hole populations in III-V and group-IV semiconductors. In particular, the absorption of circularly-polarized light in bulk Ge generates a spin-oriented population of electrons in the conduction band with a spin-polarization up to 50\%, whereas the hole spin-polarization, opposite to the electron one, can even reach values up to 83\%. In this letter, we theoretically investigate the optical injection of orbital polarization by means of circularly-polarized light in bulk Ge and we show that the latter considerably exceeds 100\% for holes and photon energies close to the direct Ge gap. These results suggest that Ge is a convenient platform for future development of orbitronics and opto-orbitronic devices.  

\end{abstract}

\maketitle
\medskip

In the past decades, a great effort has been devoted to the development of logic devices which might exploit the spin degree of freedom to boost the performances of electronic platforms,\cite{Zutic2004} based on the silicon mainstream, and add further logic functionalities to the common electronic digital devices. In this context, spin-transfer torque and spin-obit torque MRAMs undoubtedly represent a great success for spintronics, since these memory cells provide for a faster writing/reading speed and a lower power consumption with respect to the charge-based ferromagnetic MRAMs.\cite{Prenat2016} This results have been achieved by combining the transport properties of spin currents and the spin-charge interconversion properties of metals, such as Pt, with large spin-orbit coupling (SOC). At variance with spin-based storage devices, spin manipulation remains an elusive task: it has been shown that spin can be electically controlled in ferromagnet/normal metal systems, \cite{Ando2008} but the very low carrier spin lifetime has prevented up to now the realization of robust fully electrical spin-switch architectures. 

Orbitronics, pioneered in 2005, \cite{Bernevig2005} has emerged only in the last few years as a promising research field. \cite{Sala2022,Choi2023, Jo2024, Adamantopoulos2024, Santos2024, Santos2025} At variance with spintronics, which exploits the spin degree of freedom, orbitronics identifies the orbital angular momentum of electrons or holes as the state variable. Therefore,  generation, detection and, eventually, manipulation of orbital currents are crucial tasks for orbitronics. To this aim, one can leverage the orbital-Hall effect (OHE) \cite{Santos2024,Cullen2025,Veneri2025}, the orbitronic counterpart of the spin-Hall effect (SHE) \cite{Hirsch1999}, in which a flow of charge carriers generates transverse orbital currents. Indeed, literature reports suggest that an external electric field generates orbital currents in Ti \cite{Choi2023}, Cr \cite{Lyalin2023}, Ge \cite{Santos2024}, Si \cite{Bernevig2005,Matsumoto2025}, which are first converted into spin currents via orbital-to-spin conversion, once entered an adjacent ferromagnets, and then are able to exert a torque on their magnetization.

In this letter, we theoretically explore the generation of orbital angular momentum accumulation in bulk germanium by means of the optical orientation technique \cite{Lampel1968a, Allenspach1983, Bottegoni2020}. We exploit a 30-band $\mathbf{k\cdot p}$ method to calculate the bandstructure of bulk Ge across the whole Brillouin zone, then we evaluate the carrier, spin and obital injection rate as a function of the incident photon energy and finally we address the degree of spin and orbital polarization achieved for elecrons and holes in the conduction and valence band, respectively, due to the absorption of circularly polarized photons. We show that when the incident photon energy is close to the direct Ge gap, the orbital accumulation in the conduction band of Ge is less than$1\%$, whereas  it is as large as about $160$\% in the valence band, leading to a consistent orbital polarization. These results suggest that bulk Ge is a natural platform for orbitronic architectures, able to generate considerable orbital currents.
\begin{figure}[b]
	\includegraphics[width = \columnwidth]{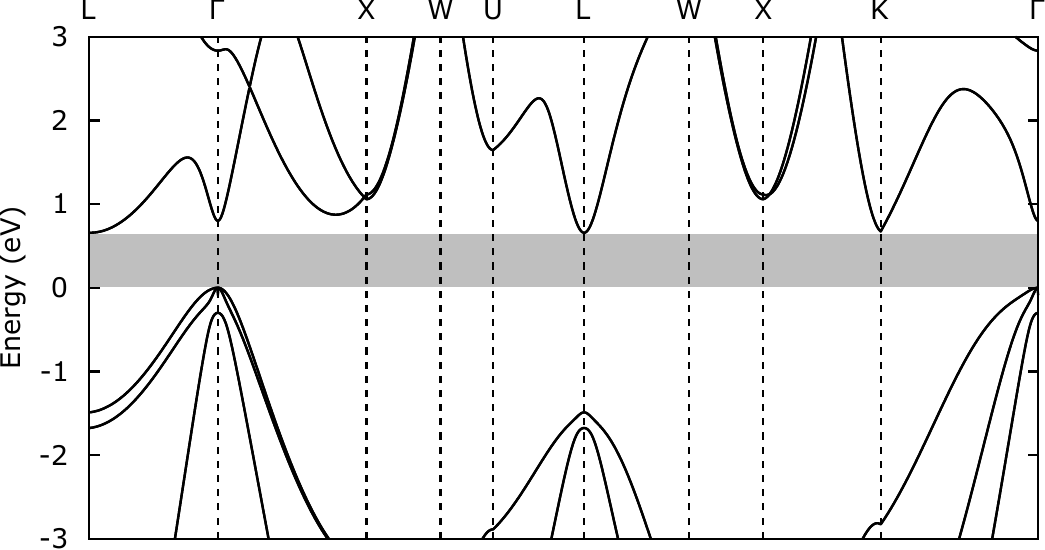}
	\caption[didascalia]{Bandstructure of bulk germanium.}
	\label{band}
\end{figure}

\begin{figure}
	\includegraphics[width = \columnwidth]{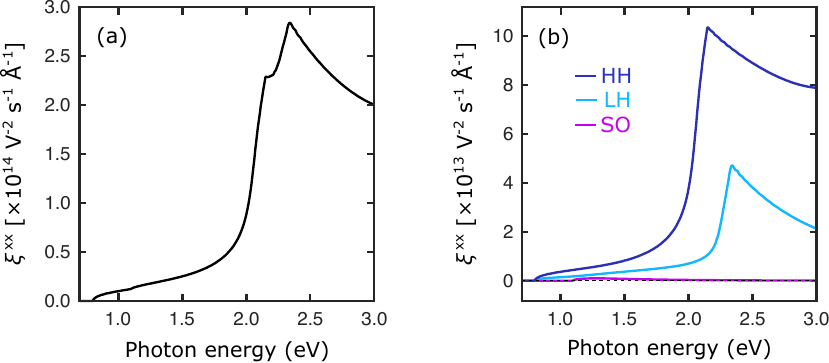}
	\caption[didascalia]{(a) Total carrier injection tensor $\xi^{xx}(\omega)$ as a function of the incident photon energy. (b) Decomposition of the total carrier injection rate from HH, LH and SO bands. }
	\label{carrinj}
\end{figure}

We have employed a $\mathbf{k \cdot p}$ model inlcluding a total of 30 states of \textit{s}-, \textit{p}- and \textit{d}- ($e_g$) type, that has already been proven effective in computing the band structure of bulk Ge throughout the entire Brillouin zone.\cite{Rideau2006}. The eigenvalues, spin-orbit splitting parameters and the matrix elements of the linear momentum are taken from Ref. \citenum{Rideau2006} and adapted to room temperature. The band structure that we obtain is reported in Fig.~\ref{band}. The carrier, spin and orbit injection rates have been calculated based on the model already developed for spin injection in Refs.\citenum{Rioux2010}, \citenum{Nastos2007} in the frame of the linear response theory. We consider a monochromatic electric field $\mathbf{E}(t)=\mathbf{E}(\omega)e^{-i\omega t}+ c.c.$, where $\omega$ is the frequency, with $\hbar\omega\geq \varepsilon_\mathrm{dg}$, being $\varepsilon_\mathrm{dg}$ the direct gap of Ge. Note that $\mathbf{E}(\omega)$ is a complex vector if it describes elliptically-polarized light. The generic component $\xi^{\alpha\beta}(\omega)$ of the carrier injection tensor can be expressed as \cite{Nastos2007}:
\begin{equation}
	\xi^{\alpha\beta}(\omega)=\frac{2\pi e^{2}}{(\hbar\omega)^{2}}\sum_{c,v}\int\frac{d\mathbf{k}}{8\pi^{3}}\hat{v}_{cv}^{\alpha\ast}(\mathbf{k})\hat{v}_{cv}^{\beta}(\mathbf{k})
	\delta[\omega_{cv}(\mathbf{k})-\omega], \label{carrinjtens}
	\end{equation}
where $e$ is the electron charge, $\alpha,\beta,\gamma$ stand for the cartesian components along the cubic axes of the crystal, the index \textit{c}(\textit{v}) indicates the summation over the states of the conduction (valence) band and $\hat{v}_{ij}^{\alpha}$ represent the matrix element of the velocity operator $\hat{v}^{\alpha}$ along $\alpha$ direction between bands $i$ and $j$ at wave vector $\mathbf{k}$: $\bra{i,\mathbf{k}}\hat{v}^{\alpha}\ket{j,\mathbf{k'}}=\hat{v}_{ij}^{\alpha}(\mathbf{k})\,\delta[\mathbf{k}-\mathbf{k}']$.\cite{Rioux2010} Integration over the entire Brillouin zone has been performed by means of the tetragonal integration method,\cite{Jeschke2016} which accurately accounts for energy conservation.\cite{Mudi2025} In cubic symmetry, the carrier injection tensor reduces to a single real-valued component: $\xi^{xx}=\xi^{yy}=\xi^{zz}$. The value of $\xi^{xx}$ is plotted in Fig. \ref{carrinj}\,\textcolor{blue}{(a)}, while the contributions of optical transitions excited from heavy hole (HH), light hole (LH) and split-off (SO) bands are reported in panel Fig. \ref{carrinj}\,\textcolor{blue}{(b)}. The carrier injection rate $\dot n$ can be then retrieved from the relation $\dot n=\xi^{xx}(\omega)E^{x}(-\omega)E^{x}(\omega)$.\cite{Najmaie2003}
\\Similarly, for the spin injection tensor, a unique independent component exists, $\zeta^{xyz}_{e(h)}=-\zeta^{xzy}_{e(h)}$, along with its cyclic permutations of the indices. It is purely imaginary, and the electron and hole contributions can be written as:
 \begin{align}
 	\zeta^{xyz}_\mathrm{e}(\omega)=\frac{\hbar}{2}\frac{\pi e^{2}}{(\hbar\omega)^{2}}\sum_{c,\bar{c},v}\int\frac{d\mathbf{k}}{8\pi^{3}}\hat{S}_{c\bar{c}}^{x}(\mathbf{k})\hat{v}_{cv}^{y\ast}(\mathbf{k})\hat{v}_{\bar{c}v}^{z}(\mathbf{k}) \notag\\
 	\times(\delta[\omega_{cv}(\mathbf{k})-\omega]+\delta[\omega_{\bar{c}v}(\mathbf{k})-\omega]),
 	\label{eq:spinelinj}
 \end{align}
 and 
 \begin{align}
 	\zeta^{xyz}_\mathrm{h}(\omega)=-\frac{\hbar}{2}\frac{\pi e^{2}}{(\hbar\omega)^{2}}\sum_{c,v,\bar{v}}\int\frac{d\mathbf{k}}{8\pi^{3}}\hat{S}_{\bar{v}v}^{x}(\mathbf{k})\hat{v}_{cv}^{y\ast}(\mathbf{k})\hat{v}_{c\bar{v}}^{z}(\mathbf{k})  \notag\\
 	\times(\delta[\omega_{cv}(\mathbf{k})-\omega]+\delta[\omega_{c\bar{v}}(\mathbf{k})-\omega]),
 	\label{eq:spinhoinj}
 \end{align}
respectively. The $\bar{c}$ ($\bar{v}$) symbols refer to a restriction on the summation related to degenerate or quasidegenerate couples of states for which $\hbar\omega_{c\bar{c}(v\bar{v})}<k_\mathrm{B}T$, where $k_\mathrm{B}$ is the Boltzmann constant, being the thermal energy set to 26 meV in our calculations. This is done to account for the coherence between states in spin-injection as discussed in Ref. \citenum{Rioux2010}. Similarly to velocity, $\hat{S}^{\alpha}_{ij}$ represent the matrix elements of the spin operator $\hat{S}^{\alpha}$ along the $\alpha$ axis between bands $i$ and $j$ at wave vector $\mathbf{k}$: $\bra{i,\mathbf{k}}\hat{S}^{\alpha}\ket{j,\mathbf{k'}}=\hat{S}_{ij}^{\alpha}(\mathbf{k})\,\delta[\mathbf{k}-\mathbf{k}']$.\cite{Rioux2010}
The spin injection rate $\dot{S}^{x}_\mathrm{e(h)}(\omega)$ can finally be calculated as $\dot{S}^{x}_\mathrm{e(h)}(\omega)=\mathrm{Im}[\zeta^{xyz}_\mathrm{e(h)}(\omega)]E^{y}(-\omega)E^{z}(\omega)$ for circularly-polarized light propagating along the $x$ direction.\cite{Najmaie2003} It should be noted that a nonzero $\dot{S}$ is obtained only if two orthogonal components of the field are phase-shifted with respect to each other, corresponding to elliptical polarization. Hence, for the same incident optical power, the injection rate is proportional to the degree of circular polarization of the light. In the following, we will not distinguish the direction of the light wavevector, since, given the cubic symmetry of the Ge crystal $\dot{S}^{x}_\mathrm{e(h)}=\dot{S}^{y}_\mathrm{e(h)}=\dot{S}^{z}_\mathrm{e(h)}$, provided that the unit vector of the spin polarization is directed along the wavevector of the circularly-polarized light.
 \begin{figure}[b]
 	\includegraphics[width = \columnwidth]{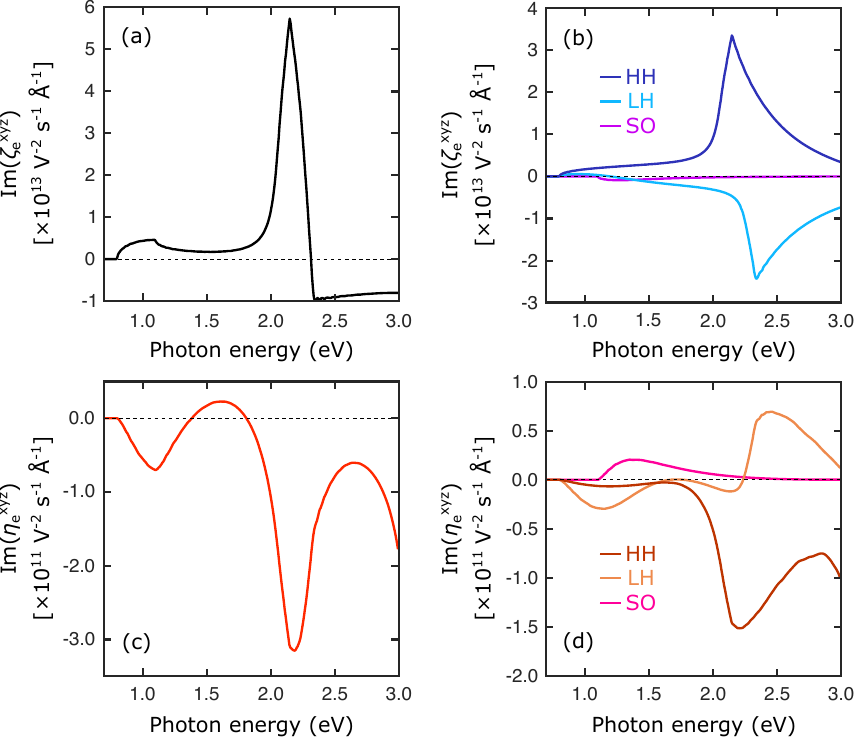}
 	\caption[didascalia]{(a) Electron spin injection pseudotensor $\zeta^{xyz}_\mathrm{e}(\omega)$ as a function of the incident photon energy. (b) Decomposition of $\zeta^{xyz}_\mathrm{e}(\omega)$ from HH, LH and SO bands. (c) Electron orbital injection pseudotensor $\eta^{xyz}_\mathrm{e}(\omega)$ as a function of the incident photon energy. (b) Decomposition of $\eta^{xyz}_\mathrm{e}(\omega)$ from HH, LH and SO bands. }
 	\label{pseudoel}
 \end{figure}
\\
The unique indipendent component of the orbital injection pseudotensor $\eta^{xyz}(\omega)$ can be similarly calculated from Eqs. \eqref{eq:spinelinj} and \eqref{eq:spinhoinj} by formally substituting $\hat{S}^{x}\rightarrow\hat{L}^{x}$, where $\hat{L}^{x}$ is the orbital angular momentum operator, and $\xi^{xyz}\rightarrow\eta^{xyz}$.
In this case, it is worth considering that the matrix elements for $s$ and $p$ orbitals of the angular momentum operator impose $\bra{s\,}\hat{L}^{\alpha}\ket{s\,}=\bra{s\,}\hat{L}^{\alpha}\ket{p,d\,}=\bra{p\,}\hat{L}^{\alpha}\ket{d\,}=0$, while $\bra{p\,}\hat{L}^{\alpha}\ket{p\,}$ result:
\begin{subequations}
\begin{equation}
\hat{L}^x_{(p)}=\hbar
\begin{pmatrix}
0 & 0 & 0 \\
0 & 0 & -i \\
0 & i & 0
\end{pmatrix}, \quad
\end{equation}
\begin{equation}
\hat{L}^y_{(p)}=\hbar
\begin{pmatrix}
0 & 0 & i \\
0 & 0 & 0 \\
-i & 0 & 0
\end{pmatrix}, \quad
\end{equation}
\begin{equation}
\hat{L}^z_{(p)}=\hbar
\begin{pmatrix}
0 & -i & 0 \\
i & 0 & 0 \\
0 & 0 & 0
\end{pmatrix}, \quad
\end{equation}
\end{subequations}
with basis states $\ket{p_x}$, $\ket{p_y}$, $\ket{p_z}$, and $\bra{d\,}\hat{L}^{\alpha}\ket{d\,}=0$ for $\ket{d_{z^2}}$, $\ket{d_{x^2-y^2}}$ orbitals. As a consequence, the orbital injection rate $\dot{L}^{x}_\mathrm{e(h)}(\omega)$ is again retrieved as ${\dot{L}^{x}_\mathrm{e(h)}(\omega)=\mathrm{Im}[\eta^{xyz}_\mathrm{e(h)}(\omega)]E^{y}(-\omega)E^{z}(\omega)}$,\cite{Rioux2010} and the same considerations of $\dot{S}$ holds also for $\dot{L}$. The imaginary components $\zeta^{xyz}_\mathrm{e}(\omega)$ and $\eta^{xyz}_\mathrm{e}(\omega)$ of the injection pseudotensor for spin and orbital angular momentum are plotted in Fig. \ref{pseudoel}\textcolor{blue}{(a,c)}, respectively, whereas the different contributions coming from transitions excited from HH, LH and SO bands are presented in panels \ref{pseudoel}\textcolor{blue}{(b,d)}. 
\begin{figure}[b]
	\includegraphics[width = \columnwidth]{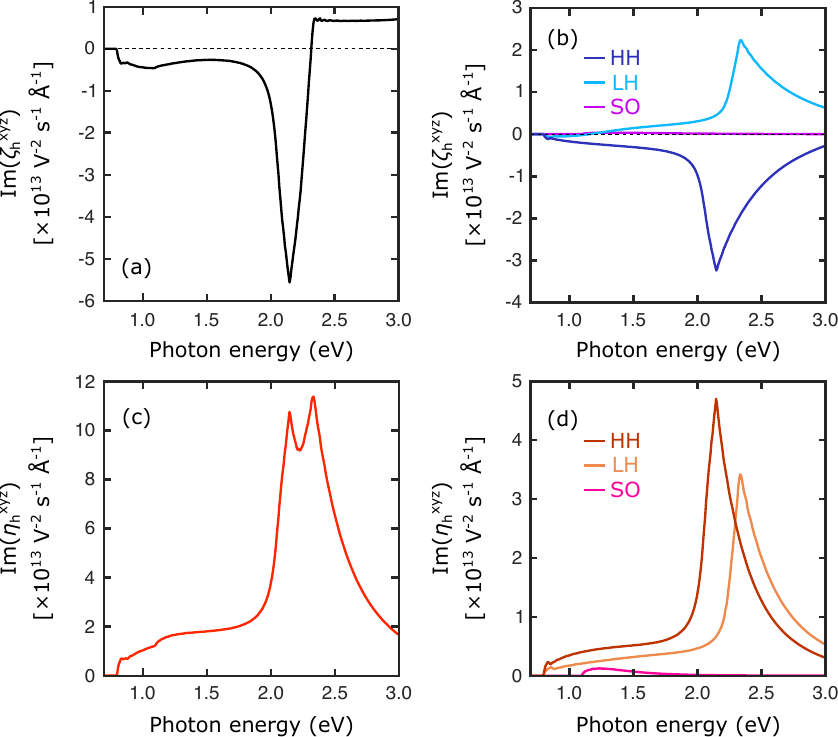}
	\caption[didascalia]{(a) Hole spin injection pseudotensor $\zeta^{xyz}_\mathrm{h}(\omega)$ as a function of the incident photon energy. (b) Decomposition of $\zeta^{xyz}_\mathrm{h}(\omega)$ from HH, LH and SO bands. (c) Hole orbital injection pseudotensor $\eta^{xyz}_\mathrm{h}(\omega)$ as a function of the incident photon energy. (b) Decomposition of $\eta^{xyz}_\mathrm{h}(\omega)$ from HH, LH and SO bands. }
	\label{pseudoho}
\end{figure}
The corresponding results for holes in the conduction band of bulk Ge are shown in Fig. \ref{pseudoho}, where panels \textcolor{blue}{(a)} and \textcolor{blue}{(c)} refer to the $\zeta^{xyz}_\mathrm{h}(\omega)$ and $\eta^{xyz}_\mathrm{h}(\omega)$ components, and the different contributions coming from HH, LH and SO states are presented in panels \ref{pseudoho}\,\textcolor{blue}{(b,d)}.
\begin{figure}[t]
	\includegraphics[width = 0.5\textwidth]{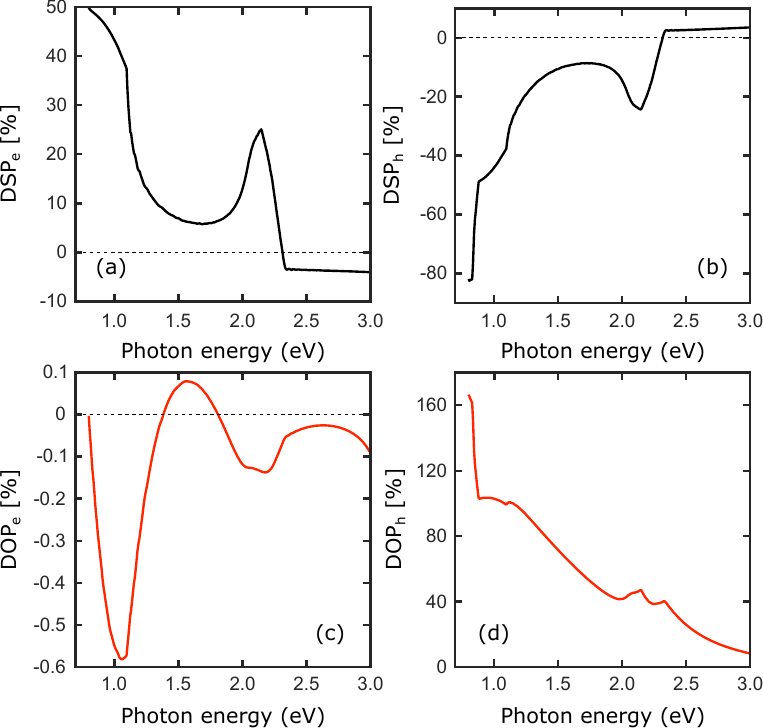}
	\caption[didascalia]{Photon energy dependence of the degree of (a) electron spin polarization DSP$_\mathrm{e}$ (b) hole spin polarization DSP$_\mathrm{h}$ (c) electron orbital polarization DOP$_\mathrm{e}$ (d)  hole orbital polarization DOP$_\mathrm{h}$.}
	\label{degree}
\end{figure}
The degree of spin polarization (DSP) and the degree of orbital polarization (DOP) for electrons and holes represents the fraction of injected carriers that have a net spin or obital-polarization. It can be defined as ${\text{DSP}_\mathrm{e(h)}=\dot{S}^{x}_{\text{e(h)}}/\dot{n}}$ and ${\text{DOP}_\mathrm{e(h)}=\dot{L}^{x}_{\text{e(h)}}/\dot{n}}$, for spin and orbit, respectively, which, for pure circularly-polarized light simplifies in ${\text{DSP}_\mathrm{e(h)}=\mathrm{Im}[\zeta^{xyz}_\mathrm{e(h)}(\omega)]/\xi^{xx}(\omega)}$ and ${\text{DOP}_\mathrm{e(h)}=\mathrm{Im}[\eta^{xyz}_\mathrm{e(h)}(\omega)]/\xi^{xx}(\omega)}$, for spin and orbit, respectively. In this case, the obtained values of DSP and DOP are reported in Fig. \ref{degree} as a function of the incident photon energy. It is worth mentioning that DOP$_{\text{h}}$ in Fig. \ref{degree} exceeds $100\%$ since it is calculated in spin $\hbar/2$ units in Eqs. \eqref{eq:spinelinj} and \eqref{eq:spinhoinj}. This allows to directly compare the amplitude of orbital angular momentum with the spin one without any change of units.
\\
It can be worth providing for a simple atomic picture to interpret the results. 
\begin{table}[b]
\footnotesize
\centering
\renewcommand{\arraystretch}{1.5}
\begin{tabular}{*{3}{c}}
\hline
\textbf{Band} & $\ket{j,m_{j}}$ & \textbf{Spherical harmonics expansion} \\
\hline
\multirow{2}*{CB} & $\ket{\sfrac{1}{2},\sfrac{1}{2}}$ & $Y_0^0\ket{\uparrow}$ \\
& $\ket{\sfrac{1}{2},-\sfrac{1}{2}}$ & $Y_0^0\ket{\downarrow}$ \\
\hline
\multirow{2}*{LH} & $\ket{\sfrac{3}{2},\sfrac{1}{2}}$ & $i\sqrt{\sfrac{1}{3}}\,Y_1^1\ket{\downarrow}+i\sqrt{\sfrac{2}{3}}\,Y_1^0\ket{\uparrow}$ \\
& $\ket{\sfrac{3}{2},-\sfrac{1}{2}}$ & $i\sqrt{\sfrac{1}{3}}\,Y_1^{-1}\ket{\uparrow}+i\sqrt{\sfrac{2}{3}}\,Y_1^0\ket{\downarrow}$ \\
\hline
\multirow{2}*{HH} & $\ket{\sfrac{3}{2},\sfrac{3}{2}}$ & $-i Y_1^1\ket{\uparrow}$ \\
& $\ket{\sfrac{3}{2},-\sfrac{3}{2}}$ & $-i Y_1^{-1}\ket{\downarrow}$ \\
\hline
\multirow{2}*{SO} & $\ket{\sfrac{1}{2},\sfrac{1}{2}}$ & $i\sqrt{\sfrac{1}{3}}\,Y_1^0\ket{\downarrow} -i\sqrt{\sfrac{2}{3}}\,Y_1^{-1}\ket{\uparrow}$ \\
& $\ket{\sfrac{1}{2},-\sfrac{1}{2}}$ & $-i\sqrt{\sfrac{1}{3}}\,Y_1^0\ket{\uparrow} +i\sqrt{\sfrac{2}{3}}\,Y_1^{1}\ket{\downarrow}$ \\
\hline
\end{tabular}
\caption{\footnotesize{Total angular momentum quantum numbers and spherical harmonics expansion of the wavefunctions for states at the $\Gamma$ point.}}
\label{tab:clebschgordan}
\end{table}
At the $\Gamma$ point of the Ge Brillouin zone, the electronic states can be written in terms of atomic $s$ and $p$ orbitals. Total angular momentum quantum numbers and spherical harmonics expansion of the wavefunctions for states at the $\Gamma$ point is reported in {Table \ref{tab:clebschgordan}}. 
The origin of the electron and hole spin-polarization DSP$_\mathrm{e(h)}$ for $\hbar\omega\approx \varepsilon_\mathrm{dg}$ stems from the fact the transition probability for electrons, promoted from HH to the conduction band, is three times larger than the LH$\rightarrow$CB transition \cite{Allenspach1983, Bottegoni2011}. Momentum conservation implies that the absorption of circularly polarized photons with, e.g., left-handed circular polarization ($\sigma^{-}$) results in a change of angular momentum $\Delta m_{j} = -1$. This allows generating 3 electrons in $m_s=\sfrac{1}{2}$ states, excited from HH, for each electron with $m_s=-\sfrac{1}{2}$, excited from the LH, thus leading to DSP$_{\text{e}}=50\%$. Following the same logic it is possible to obtain $\text{DSP}_\mathrm{h}= 83$\% in the valence band. The orbital accumulation under the optical orientation process can be justified considering that, at the $\Gamma$ point, spin-oriented electrons are promoted into the conduction band, composed of $s$-like states with $l=0$, thus leading to $\text{DOP}_\mathrm{e}=0$. Similarly, $\text{DOP}_\mathrm{e}=0$ remains quite small also for incident photon energies larger than $\varepsilon_\mathrm{dg}$ due to the small contribution of $p$-like orbitals to the electronic states out of $\Gamma$. On the contrary, if the semiconductor is illuminated with, e.g., left-handed circularly polarized light and $\hbar\omega= \varepsilon_\mathrm{dg}$, only the HH state with $m_{l}=1$ and the LH state with $m_{l}=0,1$ are involved into the optical transitions (see Table \ref{tab:clebschgordan}), with relative intensity 3:1. Specifically, for HH $\bra{\sfrac{3}{2},\sfrac{3}{2}}L^z\ket{\sfrac{3}{2},\sfrac{3}{2}}=\hbar$, and for LH $\bra{\sfrac{3}{2},\sfrac{1}{2}}L^z\ket{\sfrac{3}{2},\sfrac{1}{2}}=\sfrac{1}{3}\,\hbar$. Thus, the DOP$_\text{h}$ for photon energies resonant to the direct gap results $(3\hbar+\sfrac{1}{3}\,\hbar)/4$, leading to a DOP$_\text{h}$ value equal to $5/6=166\%$ in unit of $\hbar/2$. For photon energies larger than $\varepsilon_\mathrm{dg}$ the contribution coming from $p$-like states with $m_l=0$ increases leading to a strong decrease in $\text{DOP}_\mathrm{h}$, which nevertheless remains above 40$\%$ up to $\hbar\omega < 2.2$ eV.

In conclusion, we have theoretically explored the generation of spin and orbital accumulation when bulk Ge is illuminated with circularly polarized light. While the spin accumulation in the conduction band is quite high, the orbital accumulation is $<1\%$ due to the $s$-like nature of the electronic states around the $\Gamma$ point of the Brillouin zone. On the contrary, a very large orbital accumulation can be achieved in the valence band. This results suggest that the Ge valence band could be exploited as a highly-efficient tool for generating orbital angular momentum in semiconductors for applications in orbitronic architectures.




%


\end{document}